\documentclass[conference]{IEEEtran}
\IEEEoverridecommandlockouts
\usepackage{cite}
\usepackage{amsmath,amssymb,amsfonts}
\usepackage{algorithmic}
\usepackage{graphicx}
\usepackage{textcomp}
\usepackage{xcolor}
\usepackage{orcidlink}
\usepackage{multirow}
\usepackage{float}
\usepackage{stfloats}
\usepackage{siunitx}
\usepackage{amsmath,amsfonts}
\usepackage{algorithmic}
\usepackage{algorithm}
\usepackage{array}
\usepackage[caption=false,font=normalsize,labelfont=sf,textfont=sf]{subfig}
\usepackage{textcomp}
\usepackage{stfloats}
\usepackage{url}
\usepackage{verbatim}
\usepackage{romannum}
\usepackage[table,xcdraw]{xcolor}
\usepackage{graphicx}
\usepackage{subcaption}
\usepackage{siunitx}
\usepackage{multirow}
\usepackage{orcidlink}
\usepackage{float}
\def\BibTeX{{\rm B\kern-.05em{\sc i\kern-.025em b}\kern-.08em
    T\kern-.1667em\lower.7ex\hbox{E}\kern-.125emX}}
    
\begin{document}

\title{
Res-DPU: \textbf{Re}source-\textbf{s}hared \textbf{D}igital \textbf{P}rocessing-in-memory \textbf{U}nit for Edge-AI Workloads
}

\author{
\IEEEauthorblockN{
Mukul Lokhande\textsuperscript{\orcidlink{0009-0001-8903-5159}}, Member, IEEE, Narendra Singh Dhakad\textsuperscript{\orcidlink{0000-0003-2848-1785}}, Seema Chouhan\textsuperscript{\textdagger,\orcidlink{0009-0006-2397-5670}},\\
Akash Sankhe\textsuperscript{\textdagger,\orcidlink{0009-0006-2397-5670}},
and Santosh Kumar Vishvakarma\textsuperscript{\orcidlink{0000-0003-4223-0077}},~\IEEEmembership{Senior Member,~IEEE}
}
\thanks{\textsuperscript{\textdagger}Both authors contributed equally to this work.}
\thanks{M. Lokhande, N. S. Dhakad, S. Chouhan, and S. K. Vishvakarma are with the NSDCS Research Group, Department of Electrical Engineering, IIT Indore, Simrol 453552, India (e-mail: skvishvakarma@iiti.ac.in).}
\thanks{A. Sankhe was with the NSDCS Research Group, Department of Electrical Engineering, IIT Indore, Simrol 453552, India. He is currently with Analog Devices India Pvt. Ltd., Bangalore, India 560094.}
\thanks{This work was supported by the Special Manpower Development Program for Chip to Start-Up (SMDP-C2S), Ministry of Electronics and Information Technology (MeitY), Government of India, under Grant EE-9/2/21 - R\&D-E.}
}

\maketitle
\begin{abstract}
Processing-in-memory (PIM) has emerged as the go-to solution for addressing the von Neumann bottleneck in edge AI accelerators. However, state-of-the-art (SoTA) digital PIM approaches suffer from low compute density, primarily due to bulky bit-cells and transistor-heavy adder trees, which place limitations on macro-scalability and energy efficiency. This work introduces Res-DPU, a resource-shared digital PIM unit, with a dual-port 5T SRAM latch and shared 2T AND compute logic. This reflects per-bit multiplication cost to just 5.25T and reduced-transistor count of PIM array up to 56\% over SoTA works. Furthermore, a Transistor-Reduced 2D Interspersed Adder Tree (TRAIT) with FA-7T and PG-FA-26T, assist to reduce power consumption of adder tree up-to 21.35\% and lead to improved energy efficiency by 59\% over conventional 28T RCA designs. We propose a Cycle-controlled Iterative Approximate–Accurate Multiplication (CIA2M) approach, enabling run-time accuracy-latency trade-offs without requiring error-correction circuitry. The 16Kb REP-DPIM macro provides 0.43 TOPS throughput, and 87.22 TOPS/W energy efficiency in TSMC 65nm CMOS, and 96.85\% QoR for ResNet‑18/VGG‑16 on CIFAR‑10, including 30\% pruning. The proposed results establish a Res-DPU module for highly scalable and energy-efficient real-time edge AI accelerators.
\end{abstract}


\section{Introduction}
AI has become a powerhouse in most AIoT devices, driving capabilities such as voice assistance, face recognition, and personalised recommendations. Deep learning models are often utilised to enhance system capabilities, allowing them to recognise patterns more intelligently and make technology more intuitive. Multiply-accumulate (MAC) computations remain fundamental to the core resource demand\cite{Flex-PE}. Conventional von Neumann architectures have already reached their peak roofline performance and are limited by the transfer of data between the computational unit and storage. Processing-in-memory (PIM) has emerged as a promising solution to address the rapidly growing necessities of real-world processing with improved energy efficiency for matrix multiplication operations, by eliminating energy-hungry data transfer between memory and compute\cite{Samsung-DRAM-PIM, To-pimornot-Micron}. PIM focuses on the architectural enhancement of memory cells, incorporating computational logic (AND/NOR/XNOR) while retaining storage function\cite{MLPerf-DCIM-TSMC}. 

PIM computing is often categorized as digital and analog based on design approach. Prior PIM approaches have been significantly improved, although core challenges remain, including non-linear PVT variations, signal margin degradation, and area/energy overhead due to ADCs and Digital conversion loss \cite{AC_JSSC19}. This limits the work to 4-bit precision and limits macro applicability for advanced AI models. In contrast, digital PIM macros are more robust in nature and suffer no degradation in accuracy, with bit-precision scalability. However, they often lack flexibility in modified PIM arrays and area/energy overhead due to underutilized adder trees. 

\subsection{Prior Works \& Motivation: }

Prior works have focused on incorporating PIM logic, with storage bit-cell making the bit-cell significantly bulkier and subsequently reducing storage density significantly compared to an iso-area memory array. For instance, The bit-cell proposed in \cite{Flex-DCIM_TCASI'25} integrates 6T AND logic with RD-8T latch, Designs in \cite{6T+4T-NOR-ISSCC'21} and \cite{12T-NOR-ISSCC'22} combine CMOS 4T NOR with 6T and RD-8T latch respectively (Additionally 2T are required for input-inversion in NOR approach), and Analog Bit-cells in \cite{DASGupta_8T1C_TCASII22, XAC-MAC-BPD-tnano23, R-inmac_10T_IMC} employed 4T for XNOR based In-memory computations approach. When considered in combination with resource-heavy adder tree or ADC architectures, the storage density has dropped severely to 240 KB for 1MB L1-cache while replacing existing storage devices with advanced PIM-incorporated caches, raising the question: \textbf{To PIM or Not? In-Memory Systems}. This can be inferred in detail with several industry literature and DCIM benchmarks\cite{Samsung-DRAM-PIM, To-pimornot-Micron, MLPerf-DCIM-TSMC, AMD-Zen4c-ISSCC'22}. 

We started this work with \textbf{motivation} to address the fundamental challenge associated with compute density, i.e. extract best TOPS performance out of 1 mm\textsuperscript{2} piece of silicon. The key drivers were 1. To reduce the transistors required per bit multiplication, preferably go lower than a 6T latch to store data. 2. Incorporate time-shared logic philosophy to minimise processing area overhead. and 3. Low-resource Adder Tree, albeit with an acceptable performance-accuracy tradeoff. 

LPRE\cite{LPRE} improved compute efficiency with time-multiplexed activation function resources. Flex-DPU\cite{Flex-DCIM_TCASI'25} followed shared 6T AND logic with consecutive 8 x 8T memory cells, reducing computer overhead per bit multiplication to 0.75T. However, Flex-DCIM deployed a significantly transistor-heavy adder tree-based power-gated 26T full adder. This doesn't reduce resources compared to Conventional CMOS adder tree structures (Ripple-carry adder (RCA) approach: FA-28T), which dominates 65\% of total power consumption and 45\% of area in PIM-macro. Interspersed adder tree structures have been found major boost in this perspective. For instance, a 2D interleaved/interspersed combination of FA-28T with FA-14T\cite{6T+4T-NOR-ISSCC'21} or 7T\cite{sankhe-isqed25} has been found to be approximately three orders of magnitude energy efficient. Other approach involved approximate arithmetic to improve compute density, either with compressors\cite{DIMCA-JSSC'24} designed with Interspersed FA-12T and IFA-16T\cite{HOAA} or Lower-OR-adders\cite{LOA-TVLSI, Narendra-6T_IMC}.

\begin{table*}[!t]
\caption{Comparative benchmark between State-of-the-Art Bit Bit-cells, adapted in TSMC 65nm CMOS and VDD 1.2V}
\label{tab:compute-comp}
\renewcommand{\arraystretch}{1.35}
\resizebox{\textwidth}{!}{%
\begin{tabular}{|l|cccc|cc|c|c|c|c|c|c|}
\hline
 & \multicolumn{2}{c|}{\textbf{Standard Cell}} &  \multicolumn{1}{c|}{\textbf{TCAS-II'12\cite{Adam-TCASII'12}}} & \textbf{JSSC'11\cite{Adam-JSSC11}} & \multicolumn{1}{c|}{\textbf{ISSCC'21\cite{6T+4T-NOR-ISSCC'21}}} & \textbf{ISSCC'22\cite{12T-NOR-ISSCC'22}} & \textbf{TCAS-II'22\cite{DASGupta_8T1C_TCASII22}} & \textbf{ISQED'25\cite{sankhe-isqed25}} & \textbf{JSSC'25} & \textbf{TCAS-I'25\cite{Flex-DCIM_TCASI'25}} & \textbf{TNANO'23\cite{XAC-MAC-BPD-tnano23}} & \textbf{Proposed} \\ \hline
 
\textbf{\#T/Mult.} & \multicolumn{1}{c|}{6T} & \multicolumn{1}{c|}{8T} &  \multicolumn{1}{c|}{5T} & 9T & \multicolumn{1}{c|}{6T+4T} & 1R1W 12T & 10T1C & MUL8T & DLS-13T & 8.75T & XNOR-10T & 5.25T \\ \hline

\textbf{Area (um2)} & \multicolumn{1}{c|}{1.75} & \multicolumn{1}{c|}{2.35} &  \multicolumn{1}{c|}{1.55} & 2.8 & \multicolumn{1}{c|}{3.67} & 4.2 & 2.64 & 2.52 & 12.44 & 2.35 & 3.78 & 2.02 \\ \hline

\textbf{Power (nW)} & \multicolumn{1}{c|}{20.1} & \multicolumn{1}{c|}{23.4} &  \multicolumn{1}{c|}{17.1} & 32.3 & \multicolumn{1}{c|}{27.4} & 39.8 & 23.6 & 34 & 3.17 & 25.9 & 37.3 & 18.84 \\ \hline
\textbf{Operation} & \multicolumn{4}{c|}{Storage} & \multicolumn{2}{c|}{\begin{tabular}[c]{@{}c@{}}Storage, \\  CIM-NOR\end{tabular}} & XNOR & \begin{tabular}[c]{@{}c@{}}Storage,\\  CIM-AND\end{tabular} & Storage & Shared-AND & \begin{tabular}[c]{@{}c@{}}Storage, \\  XAC/MAC\end{tabular} & \begin{tabular}[c]{@{}c@{}}Storage, \\  Shared-AND\end{tabular} \\ \hline

\textbf{Layout Density} & \multicolumn{1}{c|}{High} & \multicolumn{1}{c|}{High} & \multicolumn{2}{c|}{Irregular} & \multicolumn{1}{c|}{Low} & Low & NR & Medium & High & High & Medium & High \\ \hline

\textbf{Read Delay (ps)} & \multicolumn{1}{c|}{106.2} & \multicolumn{1}{c|}{102.7} &  \multicolumn{1}{c|}{104.6} & 101.9 & \multicolumn{1}{c|}{127.5} & 158 & \multirow{2}{*}{1n} & 108.2 & 145 & 116 & 134 & 105.2 \\ \cline{1-7} \cline{9-13} 

\textbf{Write Delay (ps)} & \multicolumn{1}{c|}{169.6} & \multicolumn{1}{c|}{164.3} &  \multicolumn{1}{c|}{157.2} & 159.4 & \multicolumn{1}{c|}{216.7} & 282 &  & 173.2 & 277.7 & 184.5 & 237.6 & 157.8 \\ \hline
\end{tabular}}
\end{table*}

To address the aforementioned issues, this work focuses on enhancing energy efficiency and computational density. The key contributions of this study are as follows:

\begin{enumerate}
    \item Resource-shared Digital Processing-in-memory Unit (Res-DPU): To enhance the compute density of the DPIM macro, this work utilises a shared-logic philosophy for a single 2T AND multiplier logic with eight dual-port 5T SRAM latches. The cumulative number of transistors per 1b dot-product computation is 5.25T (5T storage and 2/8 = 0.25T), saving transistors up to 40\%, 48.5\% and 56\% compared to prior 8.75T AND\cite{Flex-DCIM_TCASI'25}, 10T XNOR\cite{XAC-MAC-BPD-tnano23} and 12T NOR bit-cell\cite{12T-NOR-ISSCC'22}, respectively.   
    
    \item Transistor-reduced 2D Interspersed Adder Tree: 
    This work explores the adder tree with FA7T and PG-FA26T in an Interspersed RCA fashion for parallel accumulation with a significant differential signal margin, and a lower transistor count by 21.35\% against the SoTA work. 
    
    \item Resource-efficient, Enhanced Performance Flexible DPIM (REP-DPIM) macro: 
    The proposed multi-precision macro benefits from a REP-DPIM array and a MUX-ed adder tree architecture. Res-DPU-based PIM macro shows improvements in $\times$ energy efficiency and has been evaluated across different sparse AI workloads, found to demonstrate $96.85\%$ QoR (FP32 baseline).
    
\end{enumerate}

\section{Resource-efficient, Enhanced Performance Flexible DPIM (REP-DPIM) macro}

The proposed REP-DPIM macro primarily benefits from reduced cycles in multiplication, as well as area and power reductions from the Res-DPU and the adder tree. The computations approach exploits PPA gains, albeit with a minor accuracy loss, from the CIA\textsuperscript{2}M multiplication approach, and sparsity levels. 

\begin{figure}[!t]
    \centering
    \includegraphics[width=0.85\columnwidth]{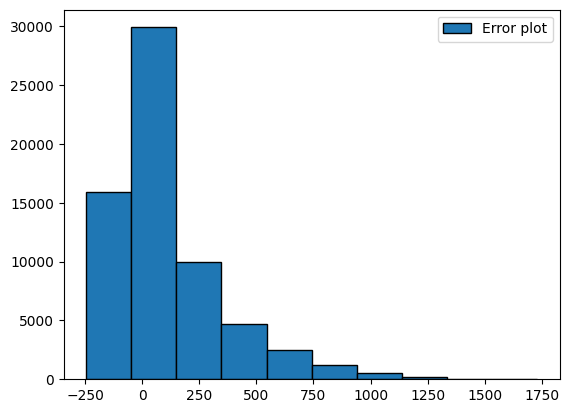}
    \caption{Visualisation of Error values (x-axis) against no of occurrences (y-axis) for the proposed CIA\textsuperscript{2}M compared to accurate INT-8b multiplication over all possible (65536) cases.}
    \label{fig:enter-label}
\end{figure}

\subsection{\textbf{C}ycle-controlled \textbf{I}terative \textbf{A}pproximate-\textbf{A}ccurate \textbf{M}ultiplication (CIA\textsuperscript{2}M) Approach}

New era AI accelerators require a comprehensive co-design methodology, integrating software-based accuracy analysis with hardware-aware architectural modelling. Software accuracy analysis becomes critical to be quantified to get the true implications of approximate quantization on real-world applications. For instance, Quant-MAC\cite{Quant-MAC} and TL-ILM\cite{ILM_TCASI21} proved to suffer heavy accuracy loss up to 4\% for VGG-16 and ResNet-20 on CIFAR-10, primarily coming from approximate truncation and 8-bit quantization. Thus, pure-approximation\cite{DIMCA-JSSC'24, Quant-MAC, ILM_TCASI21} becomes impractical for real-time applications, regardless of its hardware benefits. Extensive workload characterization has shown that the accuracy loss is often primarily due to a few sensitive and error-intolerant layers. Thus, our work focuses on Cycle-controlled Iterative Approximate-Accurate Multiplication. CIA\textsuperscript{2}M avoids the need for SoTA error-curable circuitry, which often reduces hardware benefits. CIA\textsuperscript{2}M benefits from an iterative shift-add multiplication approach in PIM-architectures, with accuracy-configurable run-time reconfiguration between accelerated approximate mode with up to 2.1-2.5\% accuracy loss in 3-cycles, while accurate mode 0.5-0.8\% accuracy loss in 4-cycles for VGG-16/ResNet-20 using CIFAR-10. The MSDF-based static segmentation beyond 4-bit has demonstrated no further implications for 8-bit multiplications\cite{App-MAC_staticseg_TETC24}. The conventional mathematical multiplication has been shown in eq. \ref{eq_mul_ac}, where P\textsubscript{acc} refers to accurate product.

\begin{equation}
\label{eq_mul_ac}
\begin{aligned} & P_{\text {acc }}=A * B=\left(A_R+2^{K a}\right) *\left(B_R+2^{K b}\right) \\
&  P_{\text {acc }}=2^{\mathrm{Ka} *} 2^{\mathrm{Kb}}+\mathrm{A}_{\mathrm{R}} * 2^{\mathrm{Kb}}+\mathrm{B}_{\mathrm{R}}{ }* 2^{\mathrm{Ka}}+\mathrm{A}_{\mathrm{R}} * \mathrm{~B}_{\mathrm{R}} \\
& P_{\text {acc }} =A * 2^{\mathrm{Kb}}+\mathrm{B}_{\mathrm{R}} * 2^{\mathrm{Ka}}+\mathrm{A}_{\mathrm{R}}{ } * \mathrm{~B}_{\mathrm{R}} \\ \end{aligned}
\end{equation}

The proposed CIA\textsuperscript{2}M computes, based on the most significant digit first, works in iterative mode similar to conventional shift-add multiplication, where the input-operand is fed bit-serially to REP-DPIM, where the weights are already stored and partial products are calculated in a cycle-wise manner. We have demonstrated the first cycle approximate product in Eq. \ref{eq_ilmop}. The process repeats on the residue of the input-operand (remainder input-bits except Leading K\textsubscript{a}), until a 3/4-bit as per static segmentation, depending on approximate or accurate computation. The proposed CIA\textsuperscript{2}M would also be applicable to previously built macros, as our approach focuses on reducing computations required without any architectural changes, thereby reducing the number of clock cycles and subsequently leading to power reduction. 

\begin{equation}
\label{eq_ilmop}
\begin{aligned}& P_{\text {app}} \quad=\operatorname{CIA^2M}(A, B)=A * 2^{\mathrm{Kb}}+B_R{ } * 2^{\mathrm{Ka}} \\
& {\text {Error = }} \operatorname{ABS}\left(P_{\text {acc }}-P_{\text {app. }}\right)=A_R * B_R \\ & \end{aligned}
\end{equation}

  \begin{figure}[!b]
    \vspace{-2mm}
    \centering
    \subfloat[]{\includegraphics[width=0.48\columnwidth, height=5.5cm]{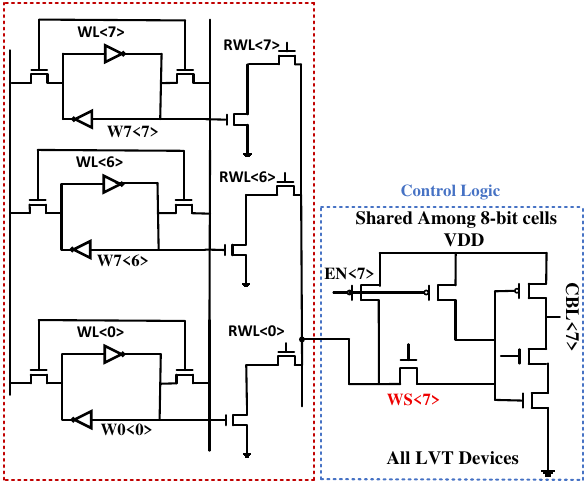}
        \label{Flex-DPU}}
    \centering
    \subfloat[]{\includegraphics[width=0.44\columnwidth]{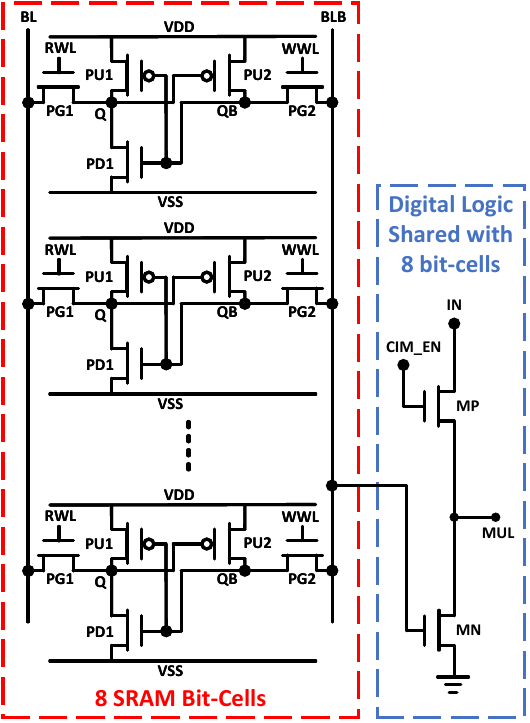}
    \label{Res-DPU}}
    \hspace{0.1\columnwidth}
    \caption{Data-path circuit comparison of Flex-DPU~\cite{Flex-DCIM_TCASI'25} and Res-DPU, highlighting reduced transistor count.}
    \vspace{-2 mm}
\end{figure}

\subsection{Resource-shared DPU (Res-DPU) and sub-bank (SBNK)}

The proposed Res-DPU primarily benefits over Flex-DPU\cite{Flex-DCIM_TCASI'25} from a Low-power Area-efficient (A5T) SRAM latch over a conventional RD8T latch, and slightly from a shared 2T AND multiplier over a prior time-shared 6T AND compute block. The proposed macro would also avoid the issues related to sense margin and brief I\textsubscript{ON}/I\textsubscript{OFF} current ratio during SRAM-latch storage due to the inherent divided bit-line structure in SBNK. Both (Res-DPU and Flex-DPU) remain iso-functional, providing word line-independent multiplication and control compute functionality with the PIM\_en signal. The detailed comparative architecture is shown in Fig. 2. The provision PIM\_en enables the selective enablement or disablement of the dot-product computation with fully reconfigurable weight precision. 

The shared-AND logic computes multiplication between input activation (WS), weight, and output product (MUL) for MAC computation, with two transistors (MP and MN). Once the PIM mode is enabled, the input (IN) is fed to the MP transistor, and the multiplication output is obtained (MUL). The multiplication functionality is validated through 100K Monte-Carlo simulations, and the standard deviation for the delay was at 11.77ps, compared to 20ps\cite{6T+4T-NOR-ISSCC'21} and 9ps\cite{Flex-DCIM_TCASI'25}. With the CIA\textsuperscript{2}M approach, we observed that only 8 × 4 sub-banks would be required for 8-b multiplication with 8x1 Res-DPUs, in contrast with an 8 × 8 sub-block of 8x1 Flex-DPU units.

\begin{figure}[!t]
    \centering
    \includegraphics[width=0.75\columnwidth]{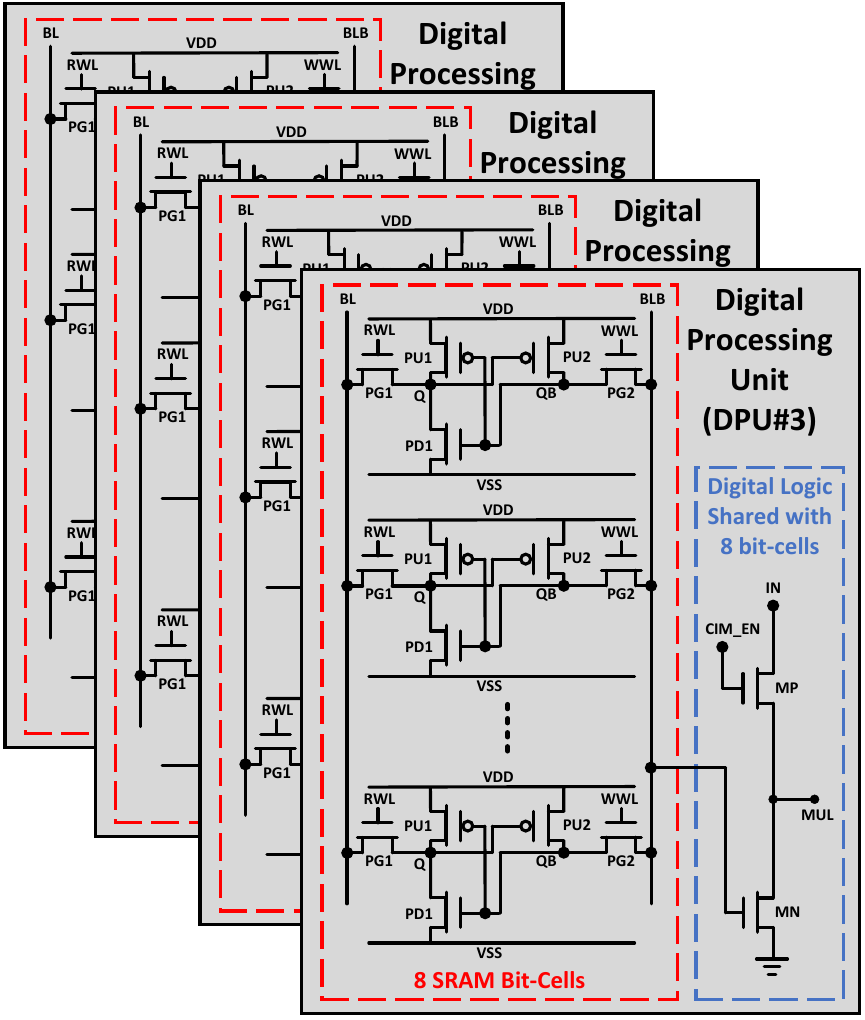}
    \caption{Proposed 8x4 sub-bank for 8-bit CIA\textsuperscript{2}M , contributes up-to two times MAC throughput compared to Flex-SBLK\cite{Flex-DCIM_TCASI'25}.}
    \label{SBNK}
\end{figure}

\subsection{Transistor-Reduced 2D Interspersed Adder Tree}
Prior works faced reduced energy efficiency, due to power consumption associated with the adder tree (up to 79\% in 4-bit macro and up to 82\% in 8-bit macro)\cite{RDCIM_TCAS-I'24}. SoTA designs replaced the traditional FA-28T-based RCA-adder tree with PG-26T \cite{Flex-DCIM_TCASI'25, RDCIM_TCAS-I'24}, resulting in an almost 35\% reduction in power consumption. This is attributed to PG-26T, depicted in Fig. \ref{26T}, a power-gated and clock-enabled FA that maintains high performance and computational accuracy; however, it contributes little to area efficiency ($<$7.14\%). The VDD1 (FA-supply) is connected to VDD during PIM mode and to VSS during storage mode. Primarily, storage mode/read data operations power-gate the adder tree, preventing leakage current. The PG-26T reduces static power consumption up to 80\%, and provides rail-to-rail swing\cite{Flex-DCIM_TCASI'25}. We incorporated 2-D interspersed adder tree structure by replacing the alternate PG-26T with prior work FA-7T\cite{sankhe-isqed25}, which contributes to reduced transistor count up to 73\% over PG-26T. The transistor schematic has been shown in Fig. \ref{7T}. Note that, even if there will be a voltage drop in the FA-7T FA, interspersed PG-26T prevents the voltage drop from being carried forward further. The detailed analysis illustrating different SoTA adder tree architectures, against No of Transistors with respective Delay(ns) and Power ($\mu$W) has been detailed in Fig. \ref{trans-delay} and \ref{trans-power}. The proposed TRAIT showcases 58.8\% and 35.7\% improvement in power and delay compared to the conventional 28T-CMOS adder tree, respectively. The work outperforms both SoTA adder tree PG-26T\cite{Flex-DCIM_TCASI'25} and 2-D interspersed (7T+28T) by 34\% and 24.3\% in Power, and 8.4\% and 48\% in Delay, respectively. The reduced power consumption and latency lead to improved energy efficiency, crucial for edge AI workloads. 

\begin{figure*}
    \centering
    \includegraphics[width=0.85\textwidth]{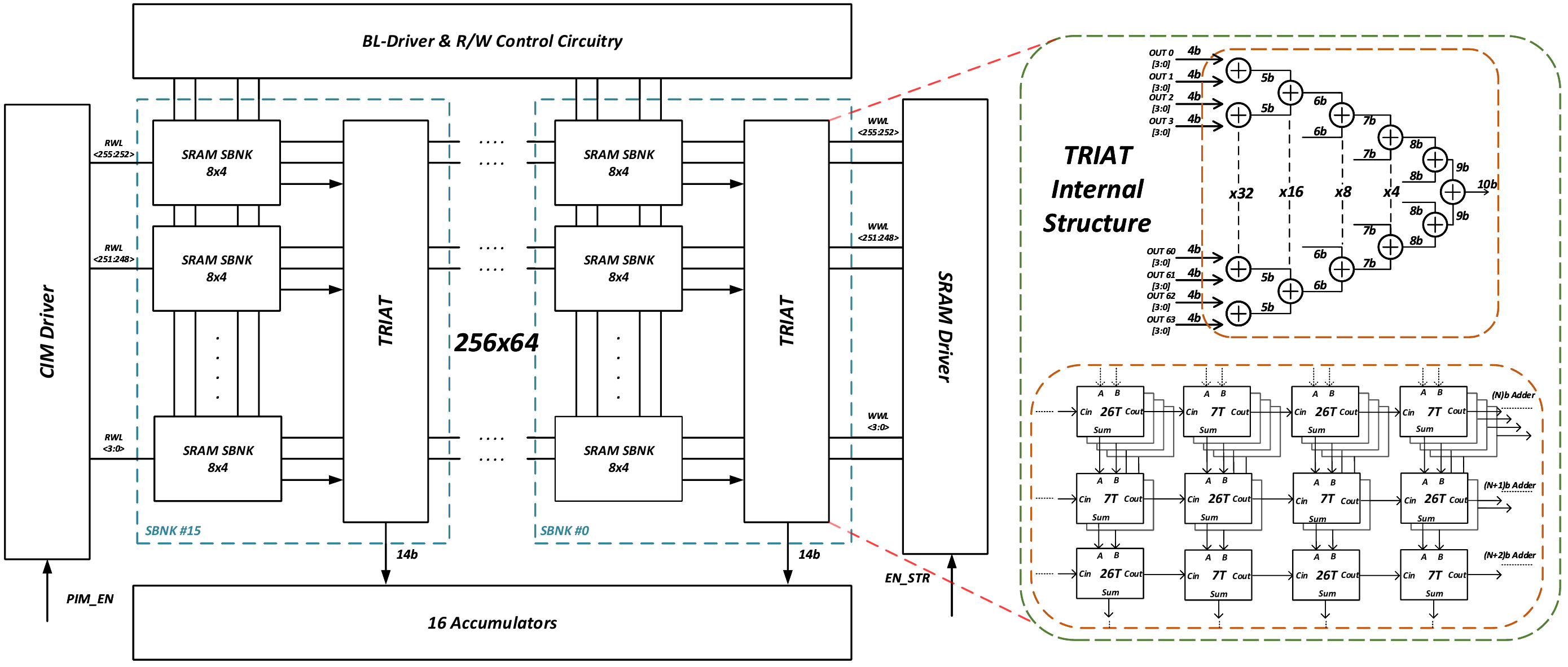}
    \caption{Illustrations of the proposed Res-DPU Macro Architecture}
    \label{macro}
\end{figure*}

\begin{figure}[!b]
    \centering
    \vspace{-4mm}
    \subfloat[]{\includegraphics[width=0.62\columnwidth]{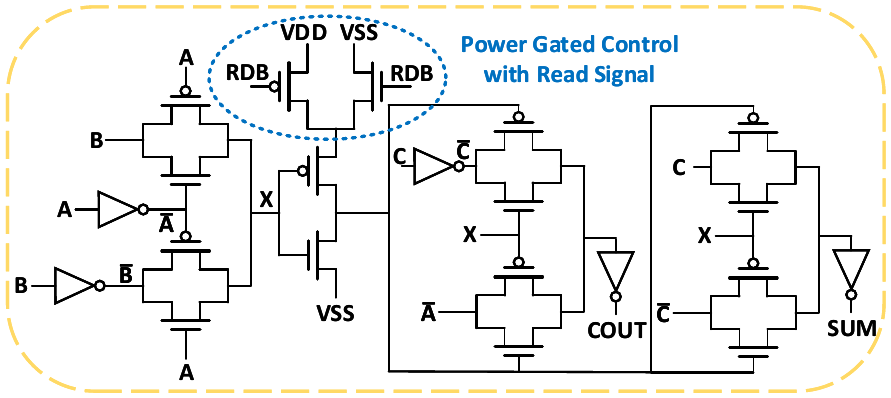}
    \label{26T}}
    \centering
    \subfloat[]{\includegraphics[width=0.36\columnwidth]{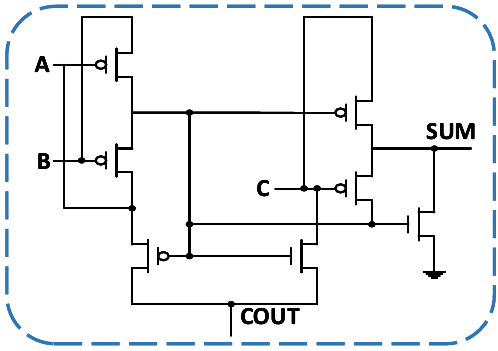}
    \label{7T}}
    \hspace{0.1\columnwidth}
    \caption{Fundamentals components of Transistor-reduced 2D-Interspersed Adder Tree (TRAIT), (a) PG-26T and (b) FA-7T.}
\end{figure}

\begin{figure}[!b]
    \vspace{-5mm}
    \centering
    \subfloat[]{\includegraphics[width=0.48\columnwidth]{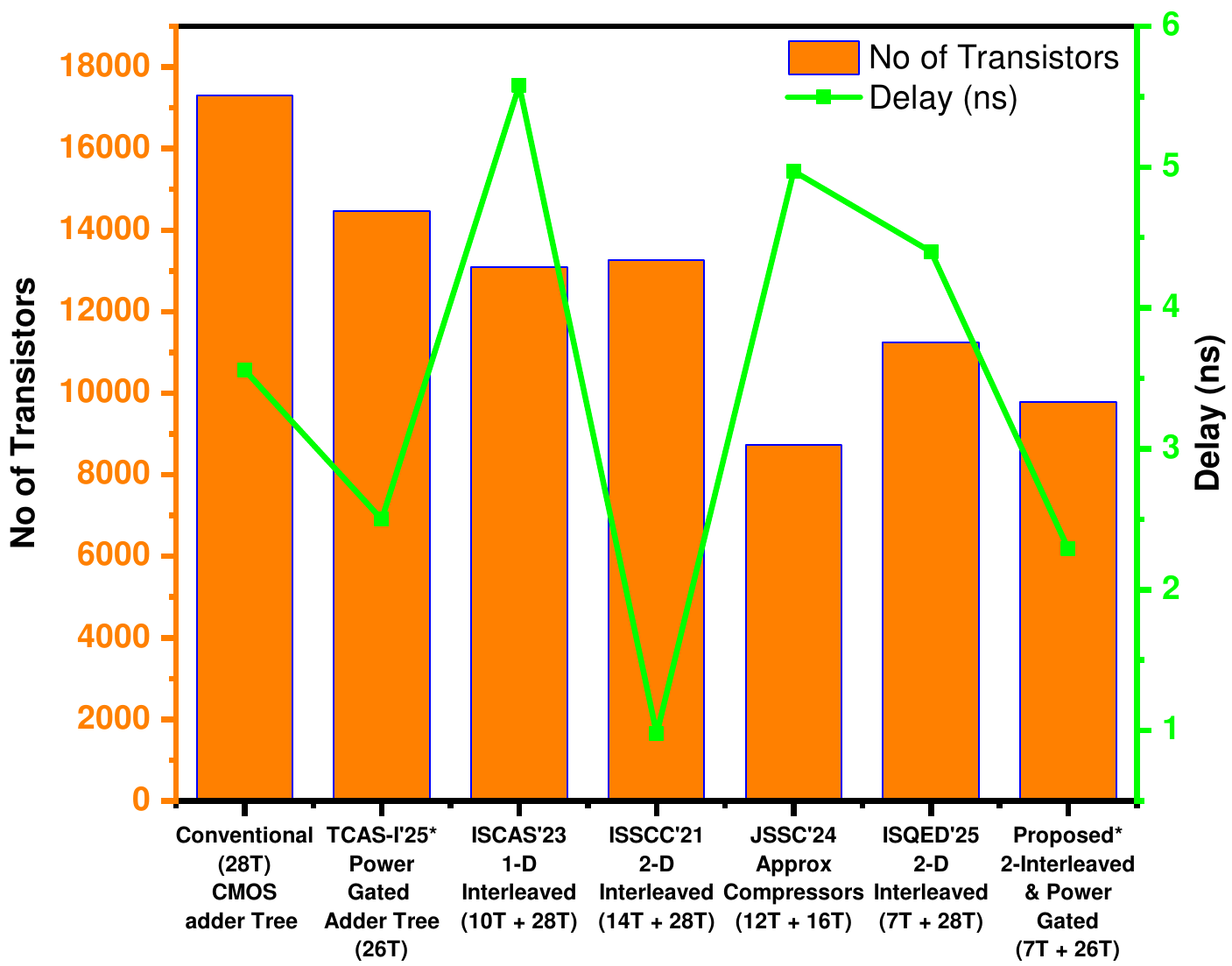}
    \label{trans-delay}}
    \centering
    \subfloat[]{\includegraphics[width=0.48\columnwidth]{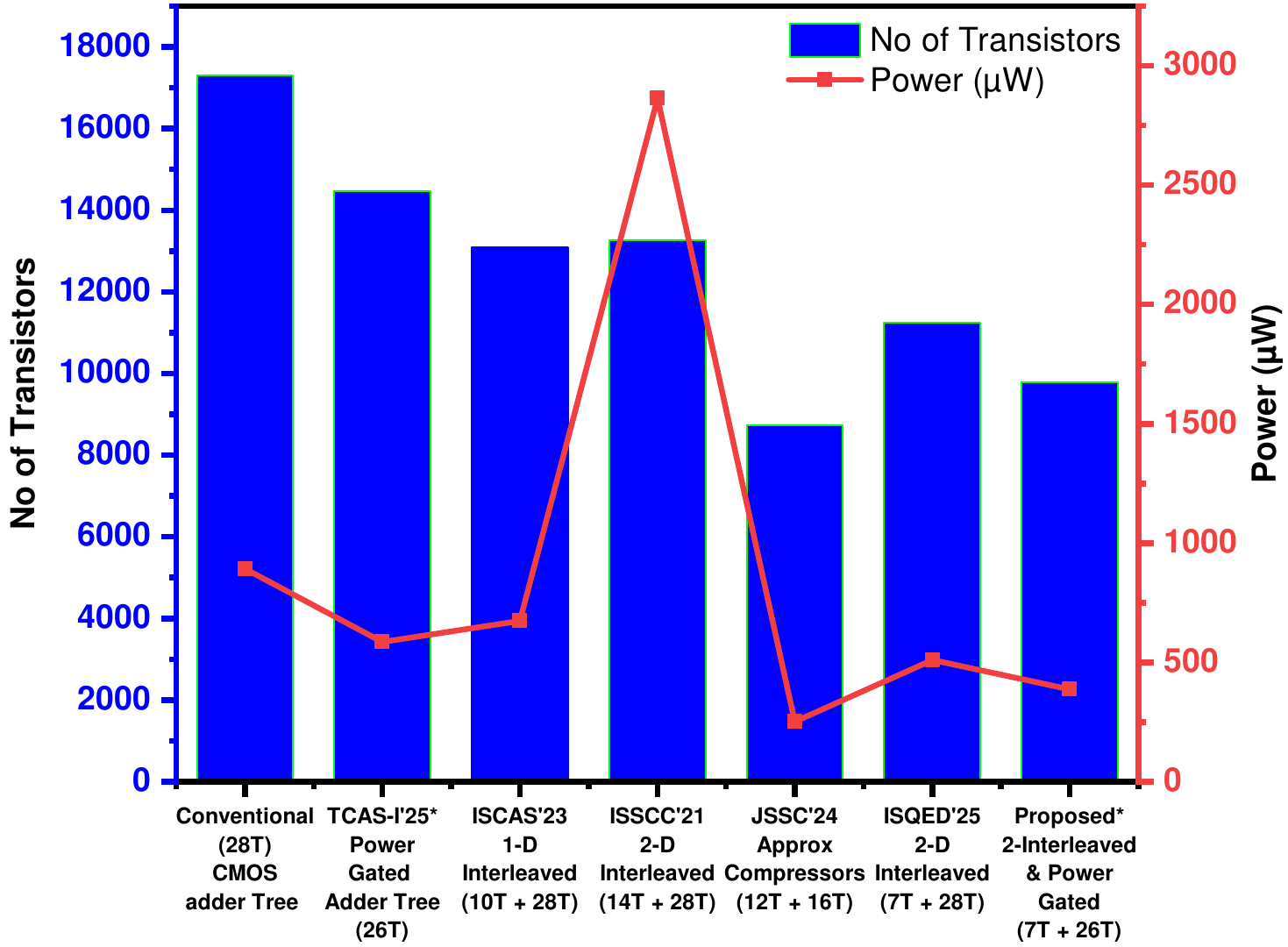}
    \label{trans-power}}
    \hspace{0.1\columnwidth}
    \caption{Performance comparison with SOTA adder tree structures, showcasing No of Transistors, (a) Delay (ns) and (b) Power ($\mu$W). }
\end{figure}

\section{Performance Analysis and Mapping Strategy}
The performance-critical workload involves MVM/MAC computations, which are best suited to be shifted with the DPIM macro. DPIM macro architecture includes an SRAM storage array, an address decoder, a controller and peripheral circuitry, an adder tree, and an accumulator. The proposed REP-DPIM macro (Fig. \ref{macro}) is 16Kb in size (256x64) divided into 8x4 SBNK architecture capable of performing multiplication operation, accompanied by TRIAT for accumulation within single clock cycle. The macro work is based on reprogramming the weight in a sequential layer format, assuming the off-chip bit-serial ReLU activation function and control engine circuitry\cite{sankhe-isqed25}. The PIM programmability is controlled by the CIM driver, and computations occur in MSDF-based bit-serial operations. The 8-bit computations with CIA\textsuperscript{2}M in 4-cycles. The structure follows Flex-DCIM and is bit-precision scalable from 1-16 bit for both input and weight. The work with 8x1 Res-DPUs utilised 9.87\% less FA compared to prior 4xN kind DCIM structures\cite{sankhe-isqed25}. 

\subsection{Hardware Analysis}
A5T SRAM-cell occupies an area of 1.55 \SI{}{\micro\meter\squared}, saving 11.43\% and 34\% CMOS area compared to the standard 6T latch and RD-8T, respectively, at a $65$\SI{}{\nano\meter} CMOS technology node. These savings contribute significantly as Flex-DPU consist of 8x1 RD-8T memory cells. The layout design of two adjacent Res-DPUs is drawn with flipped consecutive DPUs, coupled to hide the area overhead by sharing logic in the array column. Res-DPU layout design depicts dimension of 5.2 \SI{}{\micro\meter} x 6.2 \SI{}{\micro\meter} for two-adjacent 8x1 Res-DPUs compared to 5.33 \SI{}{\micro\meter} x 7.04 \SI{}{\micro\meter} for Flex-DPU. This reflects savings of up to 14\% smaller Res-DPU compared to Flex-DPU, which scales linearly from 16Kb to 64Kb for 2048 to 8192 DPUs. The Res-DPU also avoids load-parasitic issues, as 8x1 dual-port A5T bit-cells provide a resource-efficient solution and no load on the Q-node, thus avoiding bit-flip as well, unlike Flex-DPU\cite{Flex-DCIM_TCASI'25}. The pruning factor of 30\% leads to improved throughput of 0.43 TOPS and 87.22 TOPS/W energy efficiency for VGG-16/CIFAR-10 inference.

The fundamental components of TRAIT, FA-7T and PG-26T occupy an area 4.2 \SI{}{\micro\meter\squared} and 15.63 \SI{}{\micro\meter\squared} respectively compared to 22.5 \SI{}{\micro\meter\squared} in conventional FA-28T design at a 65nm CMOS technology node. The functionality was validated with 100K Monte-Carlo simulation, and it was observed that the adder tree contributes a really small just 13.6\% in the static power consumption of the total power consumption. The translation to delay(ns) and Power-consumption (mW) wrt. SoTA work's transistor count has been detailed in Fig. \ref{trans-delay} and \ref{trans-power} respectively.

The proposed REP-DPIM macro can compute 32x16, i.e., 512 1A1W dot products in a single cycle for different 32 input activations (A) and 16 weight values (W). The operating frequency is 333 MHz; thus, the throughput would be 341 TOPS. Similarly, 8A1W throughput was found to be 85.25 TOPS, as complete operation takes 4 cycles as per CIA\textsuperscript{2}M. Inversely, for fixed-8b weight precision and varied 1b/8b input precision, the throughput would be 341 and 85.25 TOPS with operations in 1 and 8 cycles, respectively. This translates to 1.66$\times$ improvement compared to \cite{Flex-DCIM_TCASI'25, XAC-MAC-BPD-tnano23}. 

The basic operation depends on the pre-charge of BLB and computation when WWL is activated, and output is taken from the MUL node. Thus, to confirm the process and temperature variation tolerance, the Res-DPUs were simulated at different process corners (FF, SS, TT) and temperatures (-10, 27, 80)\SI{}{\celsius}. The computation delay was found to be 1.296, 1.968 and 2.928 ns at FF(-10\SI{}{\celsius}), TT(27\SI{}{\celsius}) and SS(80\SI{}{\celsius}) at 333 MHz.

\begin{table*}[!t]
\caption{Performance Comparison with SoTA Low-bitwidth PIM Macros for Edge AI applications, scaled for macro-size and Tech. node}
\centering
\label{tab:perf-comp-macro}
\renewcommand{\arraystretch}{1.25}
\resizebox{0.95\textwidth}{!}{%
\begin{tabular}{|l|c|c|c|c|c|c|c|c|c|c|}
\hline
 & \textbf{Proposed} & \textbf{TCAS-I'25\cite{Flex-DCIM_TCASI'25}} & \textbf{ISQED'25\cite{sankhe-isqed25}} & \textbf{DATE'25\cite{2TAND_DATE25}} & \textbf{DARE\cite{DARE}} & \textbf{JSSC'24\cite{DIMCA-JSSC'24}} & \textbf{TCAS-I'24\cite{RDCIM_TCAS-I'24}} & \textbf{TCAS-I'23\cite{Kim_TCASI'23}} & \textbf{TNano'23\cite{XAC-MAC-BPD-tnano23}} & \textbf{JSSC'19\cite{AC_JSSC19}} \\ \hline
\textbf{Tech. (nm)} & 65 & 65 & 65 & 65 & 65 & 28 & 55 & 65 & 65 & 65 \\ \hline
\textbf{Operation} & \begin{tabular}[c]{@{}c@{}}Digital \\ Shared-AND\end{tabular} & \begin{tabular}[c]{@{}c@{}}Digital \\ Shared-AND\end{tabular} & \begin{tabular}[c]{@{}c@{}}Sparse\\ Digital AND\end{tabular} & Digital AND & Digital AND & \begin{tabular}[c]{@{}c@{}}Approximate \\ Digital XNOR\end{tabular} & CMOS NOR & Digital NMC & Digital XNOR & AMS \\ \hline
\textbf{Cell-type} & \textbf{5.25T} & 8.75T & 8T & 6T + 2T & A7T & 8T & 8T & 7T & 10T & 10T \\ \hline
\textbf{VDD (V)} & 0.7-1.2 & 1 & 1.2 & 0.6-1.2 & 0.7-1.2 & 0.45-1.1 & 1.2 & 0.8-1.1 & 1.2 & 0.8-1 \\ \hline
\textbf{Macro (Kb)} & 16 & 64 & 16 & 4 & 16 & 16 & 64 & 80 & 16 & 16 \\ \hline
\textbf{Frequency (MHz)} & 333 & 400 & 250 & 40 & 250 & 280 & 200 & 200 & 25 & 5 \\ \hline
\textbf{\begin{tabular}[c]{@{}l@{}}Area ($\mu$m\textsuperscript{2}) \\ per 1-b Mult.\end{tabular}} & \textbf{2.02} & 2.35 & 4.1 & 2.25 & 2.23 & - & 4.278 & 2.83 & 4.5 & - \\ \hline
\textbf{Input (bit)} & \textbf{1-16} & 1-8 & 4/8 & 4-8 & 1-16 & 1-4 & 4/8/12/16 & 1-16 & 4 & 6 \\ \hline
\textbf{Weight (bit)} & \textbf{1-16} & 1-8 & 1-8 & 4/8/12/16 & 4/8/12/16 & 1 & 4/8/12/16 & 1-16 & 1-4 & 1 \\ \hline
\textbf{Model} & \begin{tabular}[c]{@{}c@{}}ResNet-18,\\ VGG-16\end{tabular} & - & LeNet-5 & NA & AlexNet & CNN-8 & ResNet-18 & Inception-V4 & ResNet-20 & LeNet-5 \\ \hline
\textbf{Accuracy (\%)} & 90.1 , 89.72 & - & 99.1 & 98.5 & 98.64 & 90.41 & 94.8 & 95.3 & 98.67 & 98.3 \\ \hline
\textbf{Throughput (TOPS)} & 0.43, 1.72\textsuperscript{*} & 0.82 & 2.2 & 2.52 & 2.89 & 4.8 & - & 2.05 & 0.82 & 0.64 \\ \hline
\textbf{\begin{tabular}[c]{@{}l@{}}Energy Efficiency\\ (TOPS/W)\end{tabular}} & 87.22, 348.86\textsuperscript{*} & 249.1 & 480 & 404 & 514 & \begin{tabular}[c]{@{}c@{}}458-990,\\ 932-2219\end{tabular} & 52.22 & 63 & 273 & 51.3 \\ \hline
\end{tabular}}
\end{table*}

\subsection{Mapping Strategy \& Accuracy Evaluation}

\begin{figure}[!t]
\centering
\subfloat[]{\includegraphics[scale=0.49]{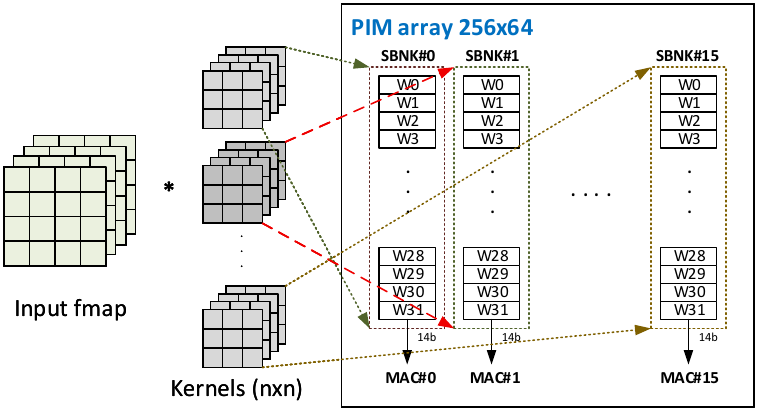}%
\vspace{-4 mm}
\label{mapping-cnn}}
\vspace{-2 mm}
\hfill
\subfloat[]{\includegraphics[scale=0.49]{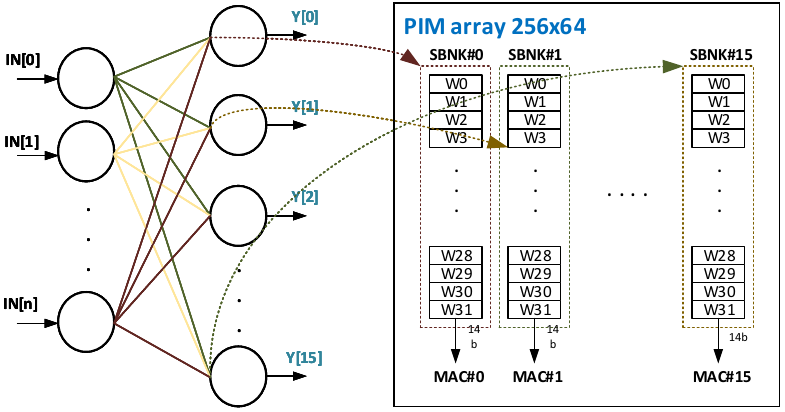}%
\label{mapping-fc}}
\caption{Visualisation for proposed REP-DPIM macro, with (a) convolution layer, (b) fully connected layer; W\textsubscript{i} represents different 8-bit weights per SBNK.}
\end{figure}

Deep neural networks comprising Convolutional and FC layers benefit from parallelism in the PIM macro. Typically, n-weight filters of size W × W × K are stored in n columns, where W represents the width of the filter and K its depth. The partial computations of the Convolution layer and FC layer are displayed in Figs. \ref{mapping-cnn} and \ref{mapping-fc}, respectively. 

We developed a parameterised software evaluation model using the $QKeras$ library with Python 3.0 on the Google Colab platform and extracted weights in CSV format for manual mapping to the REP-DPIM macro in edge-AI applications. The software-evaluation model was equipped with an error-resilient feature to assess the accuracy drop associated with quantisation, CIA\textsuperscript{2}M and 30\% pruning. The motivation behind pruning was to significantly reduce MAC computations and enhance energy efficiency with minimal accuracy loss. The hardware performance emulation model takes references from the following design parameters: DPIM array dimensions, activation and weight precision, memory banks required for layer execution, and reconfigurable hardware multiplexing. REP-DPIM array benefits from pruning due to shared digital logic and the capability to utilise independently. Our evaluation yielded application accuracies of 90.1\% and 89.72\% for ResNet-18 and VGG-16, respectively, on CIFAR-10 dataset. The FP-32 baseline shows application accuracy of 93.2\% and 92.64\% for ResNet-18 and VGG-16 using CIFAR-10, thus our approach confirms the Quality of Result (QoR) factor of 96.85\%. Our approach outperforms the 8-bit SoTA works\cite{Quant-MAC, ILM_TCASI21} with a significant margin and justifies the solution for deeper AI models at the edge.

\section{Conclusion}

This work introduces Res-DPU, a compact and energy-efficient resource-shared digital PIM macro optimised for a high-performance edge-AI accelerator.  Res-DPU benefits from a dual-port 5T SRAM latch with shared 2T AND logic, leading to a reduced per-bit multiplication cost of just 5.25T, and up to 56\% transistor savings compared to Flex-DPU. The proposed Transistor-Reduced 2D Interspersed Adder Tree (TRAIT), utilising FA-7 T and PG-FA-26 T, leads to reduced CMOS area, power, and delay by 21.35\%, 58\%, and 35.7\%, respectively, compared to conventional adder trees.
Furthermore, the proposed CIA2M method exploits the runtime accuracy-latency trade-off, without conventional error-correction circuitry, and maintains 96.85\% QoR for ResNet‑18/VGG-16 on CIFAR-10 with 30\% pruning. The proposed 16 KB REP-DPIM macro achieves a throughput of up to 0.43 TOPS and an energy efficiency of 87.22 TOPS/W. The results position Res-DPU as a strong competitor in terms of balanced performance in compute density, scalability, and energy efficiency for next-generation edge-AI SoCs. We mark the detailed exploration of compute density and the integration of the proposed macro as L3 Cache in next-generation SoCs.

\bibliographystyle{ieeetr}
\bibliography{thisbib}

\end{document}